\documentclass[structabstract]{aa}
\usepackage{graphicx}
\usepackage{txfonts}
\begin{document}

   \title{Orbital period decay of compact black hole x-ray binaries: the influence of circumbinary disks?}

   \author{Wen-Cong Chen  $^{1,3}$
          \and
      Xiang-Dong Li  $^{2,3}$
           }

   \institute{
      $^1$ School of Physics and Electrical Information, Shangqiu Normal University,
     Shangqiu 476000, China\\
     $^2$ Department of Astronomy, Nanjing University, Nanjing 210093, China;
 \\
 $^3$ Key Laboratory of Modern Astronomy and Astrophysics (Nanjing University), Ministry of Education, Nanjing 210093, China
         \\
     \email{chenwc@nju.edu.cn}
     }

\date{Received ...; accepted ...}


   \abstract
   {Recently, compact black hole X-ray binaries XTE J 1118+480 and
A0620-00 have been reported to be experiencing a fast orbital
period decay, which is two orders of magnitude higher than
expected  with gravitational wave radiation. Magnetic braking
of an Ap/Bp star has been suggested to account for the period change
when the surface magnetic field of the companion star $B_{\rm s}\ga
10^{4}$ G. However, our calculation indicates that anomalous
magnetic braking cannot significantly contribute to the
large orbital period decay rates observed in these two sources even if
$B_{\rm s}\ga 10^{4}$ G.}
   {Observations have provided evidence that
circumbinary disks around two compact black hole X-ray binaries
may exist. Our analysis shows that, for some reasonable
parameters, tidal torque between the circumbinary disk and the
binary can efficiently extract the orbital angular momentum from
the binary, and result in a large orbital period change rate.}
   {Based on the circumbinary disk model, we simulate the evolution of
XTE J 1118+480 via a stellar evolution code.}
   { Our computations are approximatively in
agreement with the observed data (the masses of two components,
 donor star radius,  orbital period, and  orbital period
derivative).}
   {The mass transfer rate and  circumbinary
disk mass are obviously far greater than the inferred
values from observations. Therefore, it seems that the circumbinary
disk is unlikely to be the main cause of the rapid orbital decay
observed in some compact black hole X-ray binaries.}

    \keywords{black hole physics -- gravitation -- stars:
individual(XTE J1118+480) --  stars: magnetic field -- X-rays:
binaries}

     \titlerunning{Orbital Period Decay of Compact Black Hole X-ray Binaries}
\authorrunning{Chen \& Li }

 \maketitle

\section{Introduction}

There are 18 stellar-mass black hole X-ray binaries (BHXBs) in our
Galaxy (\cite{remi06}; \cite{casa07}).  Half of these  have short orbital periods ($\la
0.5$ d) and light donor stars (of mass $\la 1.0~\rm
M_{\odot}$), and are classified as compact BHXBs
(\cite{lee02}; \cite{pods03}; \cite{ritt03}).
At present, the origin of compact BHXBs is still
controversial, although it is accepted that these binaries must
have experienced common envelope evolution to account for current
short orbital periods. According the theory, low-mass stars, however, are not able
to eject the envelope of the black hole progenitors during the
common envelope evolution stage (\cite{pods03}). Therefore, it was
suggested that compact BHXBs may have evolved from BHXBs with an
intermediate-mass ($\ga 2-3~\rm M_{\odot}$) donor star, i.e.,
black hole intermediate-mass X-ray binaries (BHIMXBs).
This viewpoint was supported by the observation of
CNO-processed material in XTE J1118+480 (\cite{hasw02}).

Generally, an intermediate-mass star with a radiative envelope
cannot develop large-scale magnetic fields and undergo
magnetic braking (\cite{kawa88}). During the evolution of BHIMXBs,
the orbital period should increase because  material is
transferred from the less massive secondary to the\ more massive
black hole. Short orbital periods imply that  efficient angular momentum loss mechanism must take place during the
evolution of BHIMXBs.

As an alternative scenario, BHIMXBs with
Ap/Bp stars, in which the donor stars anomalously have  strong magnetic fields of
$10^{2} - 10^{4}$ G, could experience magnetic
braking by the coupling between the magnetic field and the  winds
from the donor star driven by X-ray irradiation from the black
holes (\cite{just06}). This can cause secular decrease in the orbital periods
through tidal torques, although the efficiency of  tidal
interaction between the component with a radiative envelope and
the orbit is uncertain (\cite{devi10}). Subsequently, \cite{chen06} alternatively
proposed that a circumbinary disk around the BHXBs may offer an
efficient mechanism to extract angular momentum from the binary,
driving BHIMXBs into compact BHXBs.

Recently, \cite{gonz12} showed that the compact BHXB XTE J1118+480 is
undergoing rapid orbital decay at a rate $\dot{P}=-1.83\pm
0.66~\rm ms\,yr^{-1}$. Recently, the orbital period derivative was
updated to be $\dot{P}=-1.90\pm 0.57~\rm ms\,yr^{-1}$
(\cite{gonz14}). It was found that the orbit of A0620$-$00 is also
shrinking, though at a slightly smaller rate, i.e, $\dot{P}=-0.60\pm
0.08~\rm ms\,yr^{-1}$ (\cite{gonz14}). In Table 1, we list the
physical parameters of these two compact BHXBs.

In this work, we attempt to investigate whether the circumbinary
disk can drive the rapid decay of the orbital period measured
in these binaries. In section 2 we estimate the orbital period
derivatives in the anomalous magnetic braking  and
circumbinary disk models, and show that the former is not likely
to work for reasonable wind loss rates. In section 3, we simulate
the evolutionary sequences of XTE J1118+480, based on the circumbinary
disk model. Section 4 contains a  brief summary and discussion.

\section{The orbital decay and possible angular momentum loss mechanisms}

We consider that a BHXB  consists of a BH of mass $M_{\rm bh}$ and a donor star
of mass $M_{\rm d}$ in a circular orbit. If the mass transfer is conserved, the orbital period change
is governed by the following equation:
\begin{equation}
\frac{\dot{P}}{P}=3\frac{\dot{J}}{J}-3\frac{\dot{M}_{\rm
d}}{M_{\rm d}}\left(1-\frac{M_{\rm d}}{M_{\rm bh}}\right),
\end{equation}
where $J=2\pi a^{2}M_{\rm bh}M_{\rm d}/(M_{\rm T}P)$ is the total
orbital angular momentum of the binary, $a=(GM_{\rm
T}P^{2}/4\pi^{2})^{1/3}$ is the binary separation, $G$ is the
gravitational constant,  and $M_{\rm T}$ is the total mass of the
binary. Since $\dot{M}_{\rm d}<0$, and $M_{\rm d}\ll M_{\rm bh}$,
the second term on the right side of Equation (1) always
causes the orbital period to increase at a rate $\dot{P}\approx
-3P\dot{M}_{\rm d}/M_{\rm d}$. For a mass transfer
rate $\sim 10^{-9}~\rm M_{\odot}\,yr^{-1}$, this would indicate a
positive period derivative of $\sim 0.24~\rm ms\, yr^{-1}$
and $\sim 0.20~\rm ms\, yr^{-1}$ for XTE J1118+480 and A0620-00,
respectively. Therefore, the observed rapid orbital decay
implies that an efficient angular momentum
loss mechanism exists during the evolution BHXBs.

\subsection{Magnetic braking}

Magnetic braking originated from the coupling between the stellar
winds and the magnetic field of the donor star (\cite{verb81}). It
was proposed that the orbital decay of XTE J1118+480 may be
accounted for by magnetic braking if the donor star possesses an
anomalously strong magnetic field of $B_{\rm s} \geq10 - 20$ kG,
and the irradiation-driven wind loss rate from the donor star is
comparable to the mass transfer rate (Gonz\'{a}lez Hern\'{a}ndez et al. 2012).
According to Equation (15) in
\cite{just06}, we can obtain the orbital period derivative caused
by magnetic braking as follows:
\begin{equation}
\dot{P}_{\rm MB}=-\frac{3}{2\pi G^{1/4}}B_{\rm
s}P^{2}\psi^{1/2}\frac{\dot{M}_{\rm d}^{1/2}M^{3/2}_{\rm
T}R^{15/4}_{\rm d}}{M_{\rm bh}M^{7/4}_{\rm d}a^{7/2}},
\end{equation}
where $R_{\rm d}$ is the radius of the Ap/Bp stars, $\psi\simeq
10^{-6}c^{\rm 2}$ is a synthetic parameter taking into account the
fraction of the X-ray flux that intercepts the donor, the wind
driving energy efficiency, and the rest mass to energy conversion
efficiency.

In Figure 1, we plot the calculated rate of orbital period change
of XTE J1118+480 as a function of the mass transfer rate. One can
see that if the donor star has a strong surface magnetic field of
$10^4$ G and the mass transfer rate is  $\sim 7\times 10^{-9}~\rm
M_{\odot}\,yr^{-1}$,  magnetic braking  can indeed reproduce the
observed period change. According to  equation (9) of
\cite{king96}, the mass transfer rate driven by magnetic braking
can be written as
\begin{equation}
-\dot{M}_{\rm d}=2\times 10^{-9} \left(\frac{M_{\rm bh}}{\rm
M_{\odot}}\right)^{-2/3} \left(\frac{M_{\rm d}}{\rm
M_{\odot}}\right)^{7/3} \left(\frac{P}{3\rm hr}\right)^{5/3}\rm
M_{\odot}\,yr^{-1}.\end{equation}
For XTE J1118+480, the mass transfer rate is estimated to be
$1.6\times 10^{-11}~\rm M_{\odot}\,yr^{-1}$. Meanwhile,
\cite{wu10} found that $3 - 200$ keV peak luminosity of XTE
J1118+480 is only about 0.001 $L_{\rm Edd}$, which implied a low
mass transfer rate of $\sim10^{-10}~\rm M_{\odot}\,yr^{-1}$. In
addition, the standard magnetic braking is thought to terminate in
stars with a mass of $\la 0.3\rm M_{\odot}$, in which the donor
stars become completely convective (\cite{rapp83}; \cite{spru83}).
Therefore, some uncertainties still remain  about whether the rapid
orbital decay of XTE J1118+480 originates from magnetic braking.

\subsection{Circumbinary disk}

\cite{muno06} have detected the blackbody spectrum of compact
BHXBs A 0620$-$00 and XTE J1118+480, and found that the inferred
areas of mid-infrared 4.5-8 $\mu \rm m$ excess emission are about
two times larger than the binary orbital separations.
Recently, a detection with the Wide-field Infrared Survey
Explorer identified that XTE J1118+480 and A0620-00 are
candidate circumbinary disk systems (\cite{wang14}). These two works
strongly support the existence of circumbinary disks around
compact BHXBs. For the origin of circumbinary disks, the material
outflow during the mass transfer may be an alternative route
(\cite{dubu04}) \footnote{Circumbinary disks were also suggested to
be the remnant of fallback disks formed in the supernovae
(\cite{wang06}; \cite{cord08}).}. If a fraction $\delta$ of the transferred
mass feeds into the circumbinary disk at its inner radius $r_{\rm
i}$, tidal torque exerted on the circumbinary disk can extract the
orbital angular momentum from the BHXB. The angular momentum loss
rate can be written as (\cite{spru01})
\begin{equation}
\dot{J}_{\rm CB}=\gamma\left(\frac{2\pi
a^2}{P}\right)\delta\dot{M}_{\rm d}\left(\frac{t}{t_{\rm
vi}}\right)^{1/2},
\end{equation}
where $\gamma=\sqrt{r_{\rm i}/a}$, $t$ is the mass transfer
timescale, and $t_{\rm vi}$ is the viscous timescale at the inner edge of the
disk, $ t_{\rm vi}=2\gamma^{3}P/(3\pi\alpha\beta^{2}), $ where
$\alpha$ and $\beta$ are the viscosity parameter and the
dimensionless parameter describing the scale height at $r_{\rm i}$,
respectively (\cite{shak73}).
Adopting the typical parameters $\gamma=1.3$,
 $\alpha_{\rm SS}=0.01$, $\beta=0.03$ (\cite{chen06b}), the orbital
 period derivative caused by the circumbinary disk is given by
\begin{equation}
\dot{P}_{\rm CB}=-0.34\frac{\delta\dot{M}_{\rm d,-9}M_{\rm
T}}{M_{\rm bh}M_{\rm d}}\left(\frac{t}{142~\rm
yr}\right)^{0.5}\left(\frac{P}{1~\rm d}\right)^{0.5}~\rm
ms\,yr^{-1},
\end{equation}
where $\dot{M}_{\rm d,-9}$ is the mass transfer rate in unit of
$10^{-9}~\rm M_{\odot}\,yr^{-1}$. In Figure 2, we plot the
changing rate of the orbital period for XTE J1118+480 predicted by
the circumbinary disk model (the mass input fraction in the inner
edge of circumbinary disk is $\delta=10^{-3}$) as a function of
the mass transfer timescale. As shown in this figure, if the mass
transfer rate is $\sim 10^{-9}~\rm M_{\odot}\,yr^{-1}$,  the
evolution can lead to a change rate of $\dot{P}\approx -2.0~\rm
ms\,yr^{-1}$ at $t=10^{9}~\rm yr$. For a small
$\delta=0.000345$ (see also section 3), a long mass transfer
timescale is expected.

\begin{figure}
\centering
\includegraphics[angle=0,width=9cm]{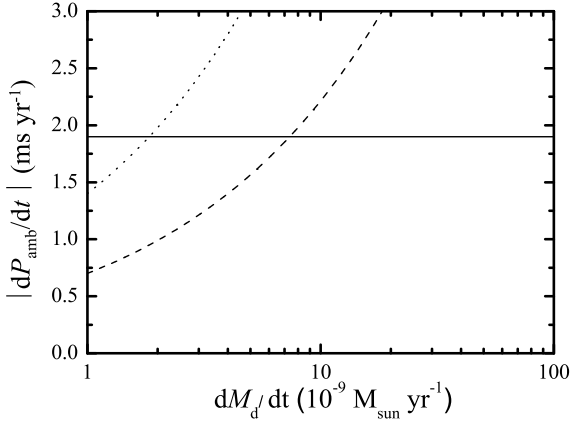}
\caption{Orbital-period change rate for XTE
J1118+480 predicted by anomalous magnetic braking as a function of
the mass transfer rate. The dashed and dotted curves correspond
to the surface magnetic fields of $10^{4}$ G and $2\times10^{4}$
G, respectively. The horizontal solid line denotes the observed
value of J1118+480.}
\label{Fig1}
\end{figure}

\section{Evolutionary simulation of XTE J1118+480}

As shown in the previous section, an overly high  mass transfer rate of
$\sim 10^{-8}~\rm M_{\odot}\,yr^{-1}$ is required to account for
the rapid orbital decay in XTE J1118+480 and A0620$-$00 if
magnetic braking works, while the circumbinary disk can address
the period derivative observed within a reasonable range of the
input parameters. Therefore, based on the circumbinary disk model
we used an updated version of the stellar evolution code developed
by Eggleton (1971,1972,1973) to simulate the evolution of XTE
J1118+480. For the donor star, we adopt an initial solar chemical
composition (X = 0.7, Y = 0.28, and Z = 0.02) and a ratio of the
mixing length to the pressure scale height of 2.0. In our
calculations, we  considered four mechanisms that are losing orbital
angular momentum, including gravitational radiation, magnetic
braking, mass loss, and the effect of circumbinary disk.
Therefore, the loss rate in the orbital angular momentum of BHXB
can be written as
\begin{equation}
\dot{J}_{\rm orb}=\dot{J}_{\rm GR}+\dot{J}_{\rm MB}+\dot{J}_{\rm
ML}+\dot{J}_{\rm CB},
\end{equation}
where four terms on the right-hand side denote angular momentum
loss rate by gravitational radiation, magnetic braking, mass loss,
and the circumbinary disk. If the mass transfer rate is higher than the Eddington accretion rate
of the BH, we assume that the redundant transferred matter is
ejected in the vicinity of the BH, forming isotropic winds and
carrying away the specific orbital angular momentum of the BH.
When the donor star mass is in the range of 0.3
-1.5 $M_\odot$, we adopt an empirical angular momentum loss
prescription for magnetic braking proposed by \cite{sill00}.

\begin{figure}
\centering
\includegraphics[angle=0,width=9cm]{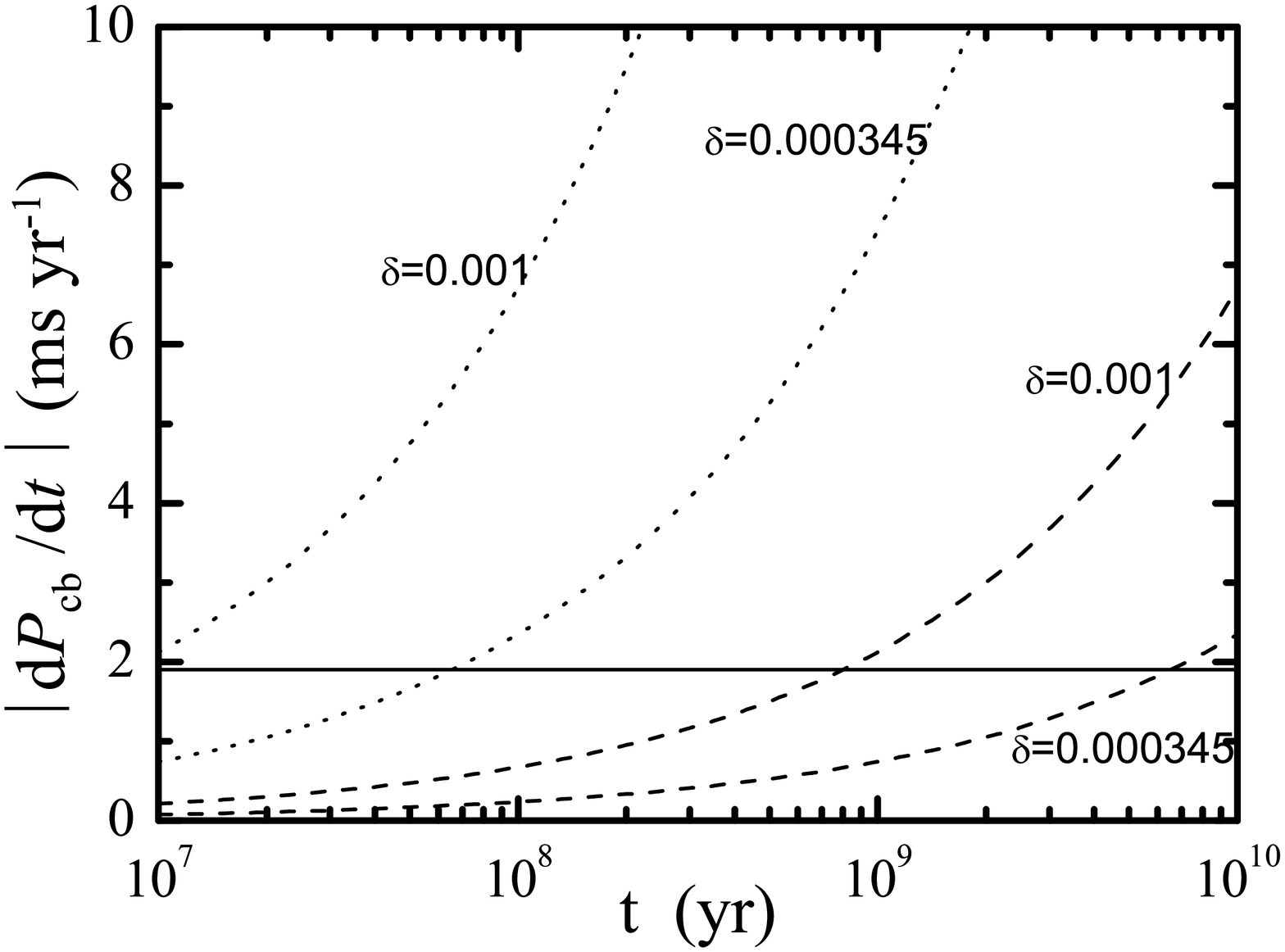}
\caption{Orbital-period change rate for XTE
J1118+480 caused by a circumbinary disk as a function of the mass
transfer timescale. The dashed and dotted curves correspond to
the mass transfer rates of $10^{-9}$ and $10^{-8}~\rm
M_{\odot}\,yr^{-1}$, respectively.  The horizontal solid line
denotes the observed value of J1118+480.}
\label{Fig1}
\end{figure}

To fit the observed parameters ( mass and radius of the donor star
and  orbital period) of XTE J1118+480, we calculated the
detailed evolutionary sequences of a BHXB with an initial orbital
period of 0.9 d, which consist of a $5.0 M_\odot$ BH and a $3.0
M_\odot$ donor star. The calculation shows that if the fraction of
the transferred mass feeds into the circumbinary disk $\delta$ is
greater than 0.00035, the mass transfer becomes dynamically
unstable. If we take $\delta=0.000345$, the calculated evolution
of the BHXB is in agreement with the observations. For the other result
$\delta\leq0.00034$, the efficiency of circumbinary disk the
extracting angular momentum is relatively low and the final orbital
period of the binary is obviously higher than the observed value.
The detailed results are summarized in Table 2. One can see that
a high $\delta$ result in a high angular momentum loss rate, a
rapid mass transfer, and a compact orbit.

\begin{table*}
 \centering
 \begin{minipage}{100mm}
\caption{Observed parameters of XTE J1118+480 and
A0620-00\label{tbl-1}}
\begin{tabular}{@{}llllll@{}}
\hline\hline
Sources & $M_{\rm bh}$   &  $ M_{\rm d}$   & $R_{\rm d}$   &$P_{\rm orb}$&$\dot{P}_{\rm orb}$\\
        & ($M_{\odot}$)  &  ($M_{\odot}$)  & ($R_{\odot}$) &   ( d )       &($\rm ms\,yr^{-1}$)\\
\hline
J1118+480 &$7.46^{+0.34}_{-0.69}$&$0.18\pm0.06$&$0.34\pm0.05$&0.169934&$-1.90\pm0.57$\\
references &                 1   &  1         & 1           &1,2,3  &1\\
 \hline
A0620-00     &$6.61^{+0.23}_{-0.17}$&$0.40\pm0.01$&$0.67\pm0.02$&0.3230142&$-0.60\pm0.08$\\
references &   1     &   1  & 1,4& 1,5&1\\
 \hline
\end{tabular}
\end{minipage}
\\References. (1)\cite{gonz14}; (2)\cite{torr04}; (3)\cite{gonz12}; (4)\cite{gonz11}; (5)\cite{mccl86}.\\
\end{table*}

In Figure 3 and 4, we show the detailed evolutionary sequences of
a BHXB with an initial orbital period of 0.9 d, which consist of a
$5.0 M_\odot$ BH and a $3.0 M_\odot$  donor star. In our calculations,
we assume that a relatively small fraction ($\delta=0.000345$) of
the transferred material forms a circumbinary disk. At the age of
$t\simeq 157.2 \rm Myr$, rapid mass transfer initiates when the
donor star fills its Roche lobe. The H-rich material is
transferred onto the BH at a rate of $\sim 10^{-8}\,\rm
M_{\odot}yr^{-1}$, which is lower than the Eddington accretion rate of the
BH. Because the material is transferred from the less massive
donor star to the more massive BH, the orbital period increases to
$~1.4$ d after $\sim 70$ Myr mass transfer. Along with the mass
transfer, the tidal torque exerted by the circumbinary disk
gradually enhances. When $\dot{P}_{\rm CB}$ = $\dot{P}_{\rm MT}$,
the orbital period of BHXB begins to decrease and the mass
transfer enters a ``plateau" phase of $\sim 400$ Myr at a rate
$\sim 10^{-9}\,\rm M_{\odot} yr^{-1}$. Once the donor star ascends
the giant branch, the mass transfer rate sharply increases to
$3\times10^{-8}\,\rm M_{\odot}yr^{-1}$, and the orbital period
rapidly decreases. At the age of $t\simeq 812.5 \rm Myr$, the
orbital period decreases to 0.17 d, and the BH and donor masses
become $7.81\rm M_{\odot}$ and $0.185\rm M_{\odot}$,
respectively. Losing most material, the radius of the donor star
decreases to be $0.336~\rm R_\odot$. The tidal torque caused by
the circumbinary disk takes away the orbital angular momentum at a
rate $-2.76~\rm ms\,yr^{-1}$, which is in general agreement with the
observation. Our simulated results indicate that the circumbinary
disk provide an alternative evolutionary channel to form compact
BHXBs.

Recently, \cite{frag09} have constrained the progenitor
properties of XTE J1118+480, and concluded that the donor star
mass is $1.0 -1.6 \rm M_{\odot}$, the BH mass is $6 -10 \rm
M_{\odot}$, and the orbital period is 0.5 - 0.8 d at the beginning
of Roche lobe overflow. Therefore, we also calculated the
evolutionary sequences a BHXB with a $6.0 M_\odot$ BH and a $1.5
M_\odot$  donor star, and an initial orbital period of 0.84 d. Similar to
the donor star of $3.0 M_\odot$, we also present the simulated result
under different input parameters $\delta$ in Table 3. One can see  that
the calculated results, when $\delta=0.00018,$ can fit the
observations. Generally, the mass transfer rates are lower for
lower donor masses, at a rate of $\sim 10^{-10} - 10^{-9}\,\rm
M_{\odot} yr^{-1}$. In the final evolutionary stage, the mass
transfer rate reached a peak of $\sim 10^{-8} - 10^{-7}\,\rm
M_{\odot} yr^{-1}$. At the age of $t\simeq 2763 \rm Myr$, the
system evolved into a compact BHXB with an orbital period of 0.17
d, which includes a $7.31~ \rm M_\odot$ BH and a $0.182~ \rm
M_\odot$ donor star. At this moment, the radius of the donor star
is $0.336~ \rm R_\odot$ and the orbital period derivative is
$-2.19~\rm ms\,yr^{-1}$.

\begin{figure}
\centering
\includegraphics[angle=0,width=9cm]{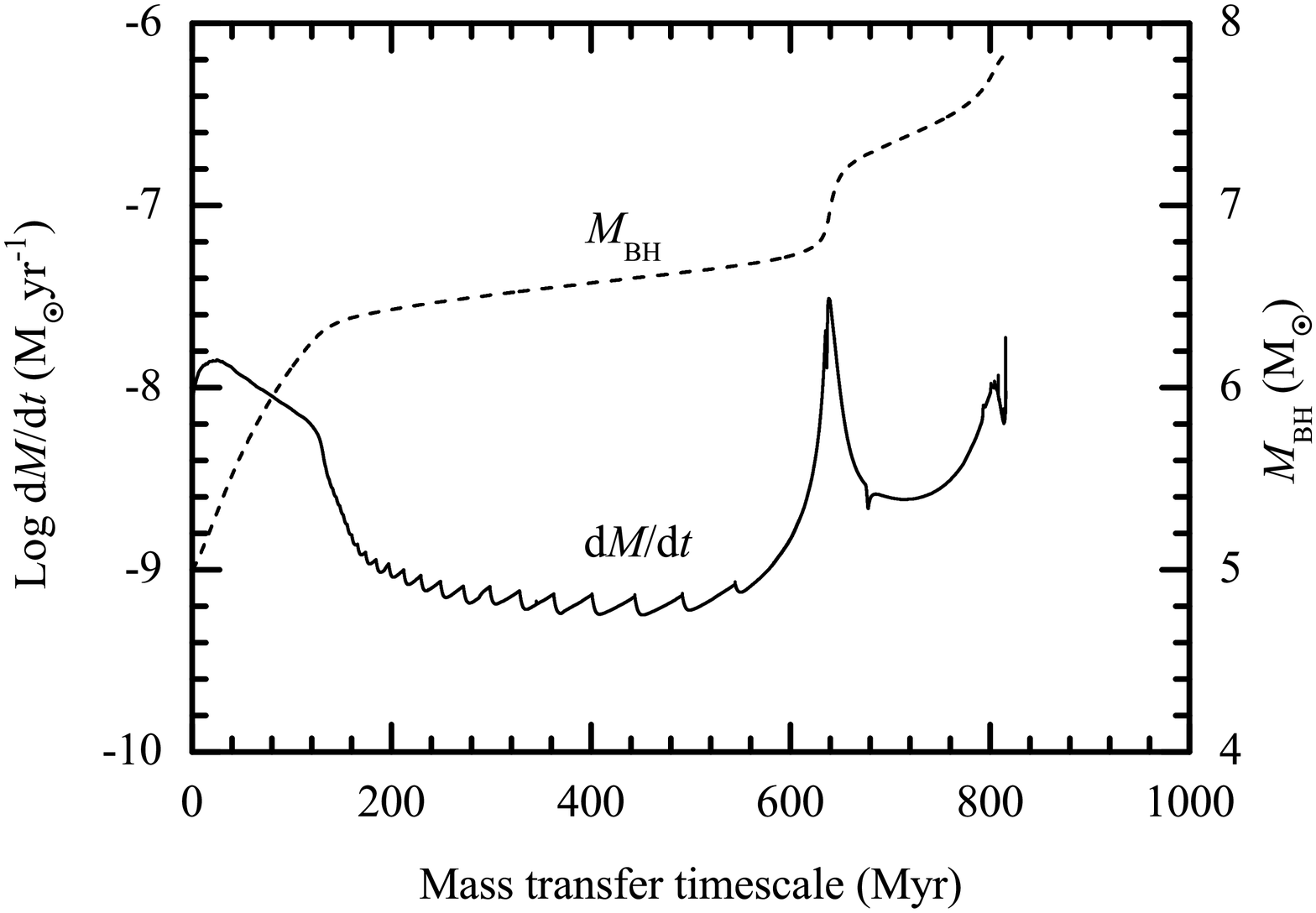}
\includegraphics[angle=0,width=9cm]{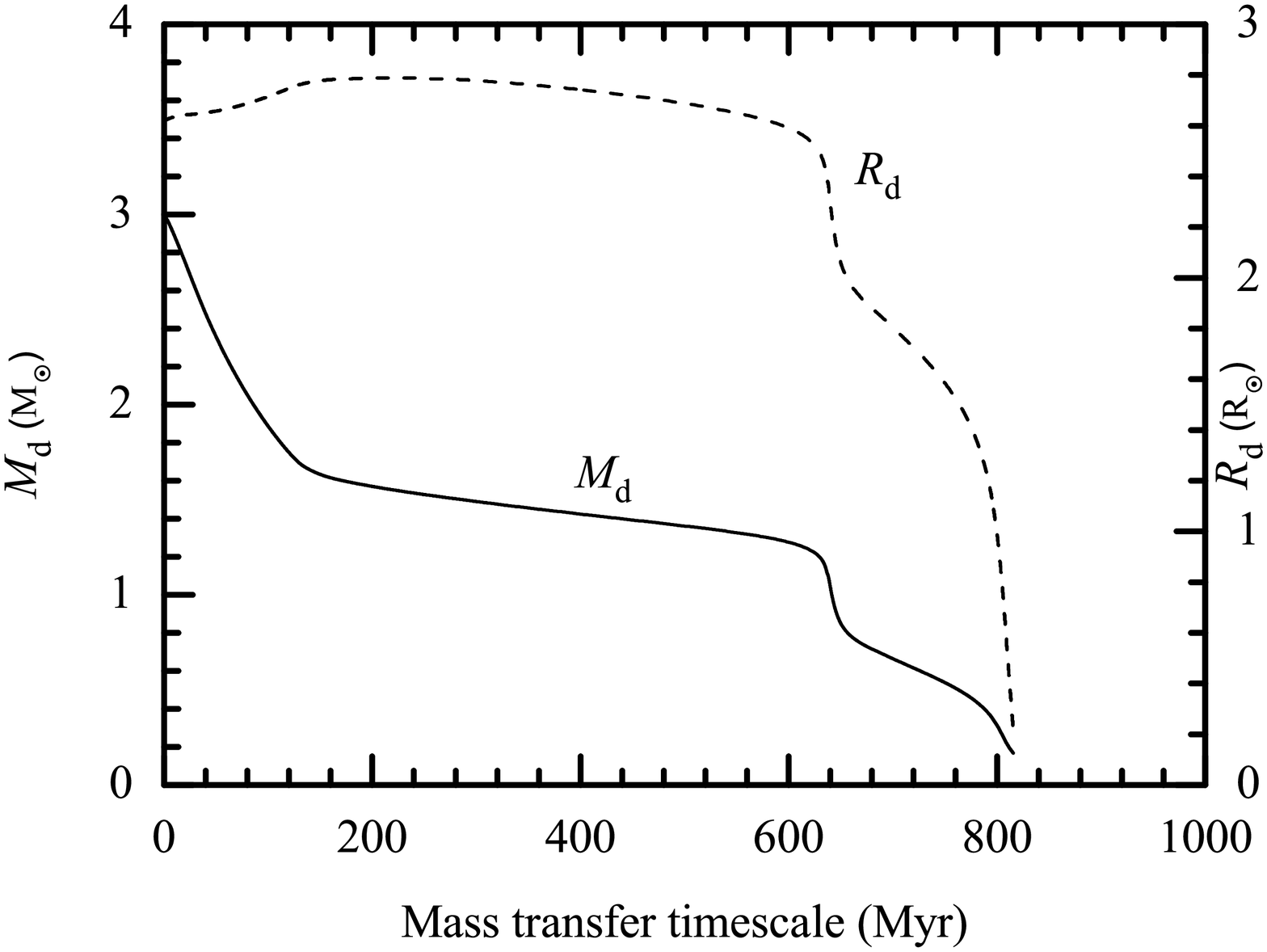}
\caption{Evolutionary tracks of BHXB
with an initial orbital period of 0.9 d, which consists of a
donor star of $M_{\rm d} = 3.0~ \rm M_\odot$ and a black hole of
$M_{\rm bh} = 5.0~ \rm M_\odot$. The solid and dashed curves
represent the evolution of the mass transfer rate, and the BH mass, in
the top panel, and the donor star mass and the donor star radius in
the bottom panel, respectively. } \label{Fig1}
\end{figure}

\begin{figure} \centering
\includegraphics[angle=0,width=9cm]{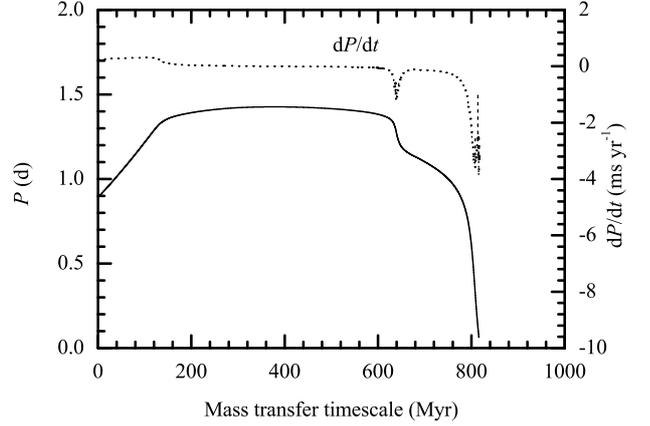}
\caption{Evolutionary tracks of BHXB with an
initial orbital period of 0.9 d, which consist of a donor star
of $M_{\rm d} = 3.0~ \rm M_\odot$ and a black hole of $M_{\rm bh}
= 5.0~ \rm M_\odot$. The solid and dotted curves represent the
evolution of the orbital period and orbital period derivative,
respectively. } \label{Fig1}
\end{figure}

\section{Discussion and summary}
In this work, we have investigated the origin of rapid orbital
decay in two compact BHXBs. Our analysis shows that even
if the donor stars of BHXBs are Ap/Bp stars with a strong magnetic
fields of 10 - 20 kG, anomalous magnetic braking scenario still
needs to solve two key problems: the mass transfer rate and
the cutoff of magnetic braking below $0.3~ \rm M_\odot$. Some
observations indicate that circumbinary disks around compact BHXBs
XTE J1118+480 and A0620-00 may exist. It is a  coincidence that rapid
orbital decay have been discovered in two compact BHXBs. If the
mass transfer timescale is $10^{9}$ yr and the mass transfer rate
is $\sim 10^{-9}~\rm M_{\odot} yr^{-1}$, the period derivative
caused by circumbinary disk with $\delta=0.001$ can account for
the observation.

\begin{table}
\begin{center}
\centering \caption{The calculated result of BHXBs with a donor
star of 3 $M_{\odot}$ under different input parameters $\delta$
\label{tbl-1}}
\begin{tabular}{cccccc}
\hline\hline
Item &     $\delta$&$M_{\rm bh}$   &  $ M_{\rm d}$     &$P_{\rm orb}$&$\dot{P}_{\rm orb}$\\
       &    & ($M_{\odot}$)  &  ($M_{\odot}$) &   ( d )    &($\rm ms\,yr^{-1}$)   \\
\hline
1$^{*}$  & 0.0002  & 7.70 & 0.29  &48.97& 1.63     \\
2$^{*}$  & 0.0003  & 7.77 & 0.23  &9.58 & 0.22 \\
3  & 0.00034  & 7.82 & 0.18  &0.54 & -0.12    \\
4  & 0.000345  & 7.82 & 0.18  &0.17 & -2.76
     \\
5  & 0.00035  & 7.82 & 0.18  &0.15 & -2.30  \\
 \hline
\end{tabular}
     \item * When the binary system has already evolved into a detached system, the donor star mass is still greater than 0.18 $M_{\odot}$.\\
\end{center}
\end{table}

Based on the circumbinary disk model, we  performed numerical
calculations for the evolution of BHIMXB with 3.0 $\rm M_{\odot}$
and 1.5 $\rm M_{\odot}$ donor star, and an initial orbital period
of 0.9 d, and 0.84 d, respectively. Our simulated results,
including the black hole mass,  donor star mass,  donor star
radius, orbital period, and  orbital period derivative, are
approximately in agreement with the observed parameters of XTE
J1118+480. It seems that the circumbinary disk can be an alternative
scenario to account for the rapid orbital decay observed
in some compact BHXBs. However, our simulations
show that the mass transfer rate is about $\sim 10^{-8}~\rm
M_{\odot} yr^{-1}$ at the current stage of XTE J1118+480, which is
obviously two orders of magnitude higher than the inferred value
from observed luminosity (Wu et al. 2010). Furthermore, there is a significant discrepancy
between our simulated mass and the observed mass of the
circumbinary disk. If the material of the circumbinary disk does
not decrease during the evolution, the mass of the circumbinary disk can be estimated to
be $\sim 10^{-9}~ {\rm M_{\odot}\,yr^{-1}}\times10^{9}~{\rm
yr}\times 0.000345 ~(0.00018)=0.000345 ~(0.00018)~{\rm
M_{\odot}}$. This mass are five to six orders of magnitude higher than the
inferred mass ($\sim10^{-9}\,M_{\odot}$) of the circumbinary disk
around XTE J1118+480 (Muno \& Mauerhan 2006). Specially,
\cite{gall07} suggested that nonthermal synchrotron emission from
a jet could account for a significant fraction (or all) of the
measured excess mid-IR emission. Therefore, the circumbinary disk
mass may be even lower. The circumbinary disk should not be the main
mechanism causing the rapid orbital decay observed in some
compact black hole X-ray binaries because the mass transfer rate and circumbinary
disk mass are obviously far greater than the inferred values
from observations(Gonz\'{a}lez Hern\'{a}ndez et al. 2012).

\begin{table}
\begin{center}
\centering \caption{The calculated result of BHXBs with a donor
star of 1.5 $M_{\odot}$ under different input parameters $\delta$
\label{tbl-1}}
\begin{tabular}{cccccc}
\hline\hline
Item &     $\delta$&$M_{\rm bh}$   &  $ M_{\rm d}$     &$P_{\rm orb}$&$\dot{P}_{\rm orb}$\\
       &    & ($M_{\odot}$)  &  ($M_{\odot}$) &   ( d )    &($\rm ms\,yr^{-1}$)   \\
\hline
1$^{a}$  & 0.0001  & 7.26& 0.25  &12.17& 0.03     \\
2$^{a}$  & 0.00015  & 7.31 & 0.21  &2.19 & 0.00 \\
3  & 0.00017  & 7.34 & 0.18  &0.59 & -0.01    \\
4  & 0.00018  & 7.33 & 0.18  &0.17 & -2.19
     \\
5$^{b}$  & 0.0002  & - & -  &- & -  \\
 \hline
\end{tabular}
     \item a When the binary system has already evolved into a detached system, the donor star mass is still greater than 0.18 $M_{\odot}$.\\
         \item b The mass transfer is dynamically unstable.\\
\end{center}
\end{table}

\begin{figure}
\centering
\includegraphics[angle=0,width=9cm]{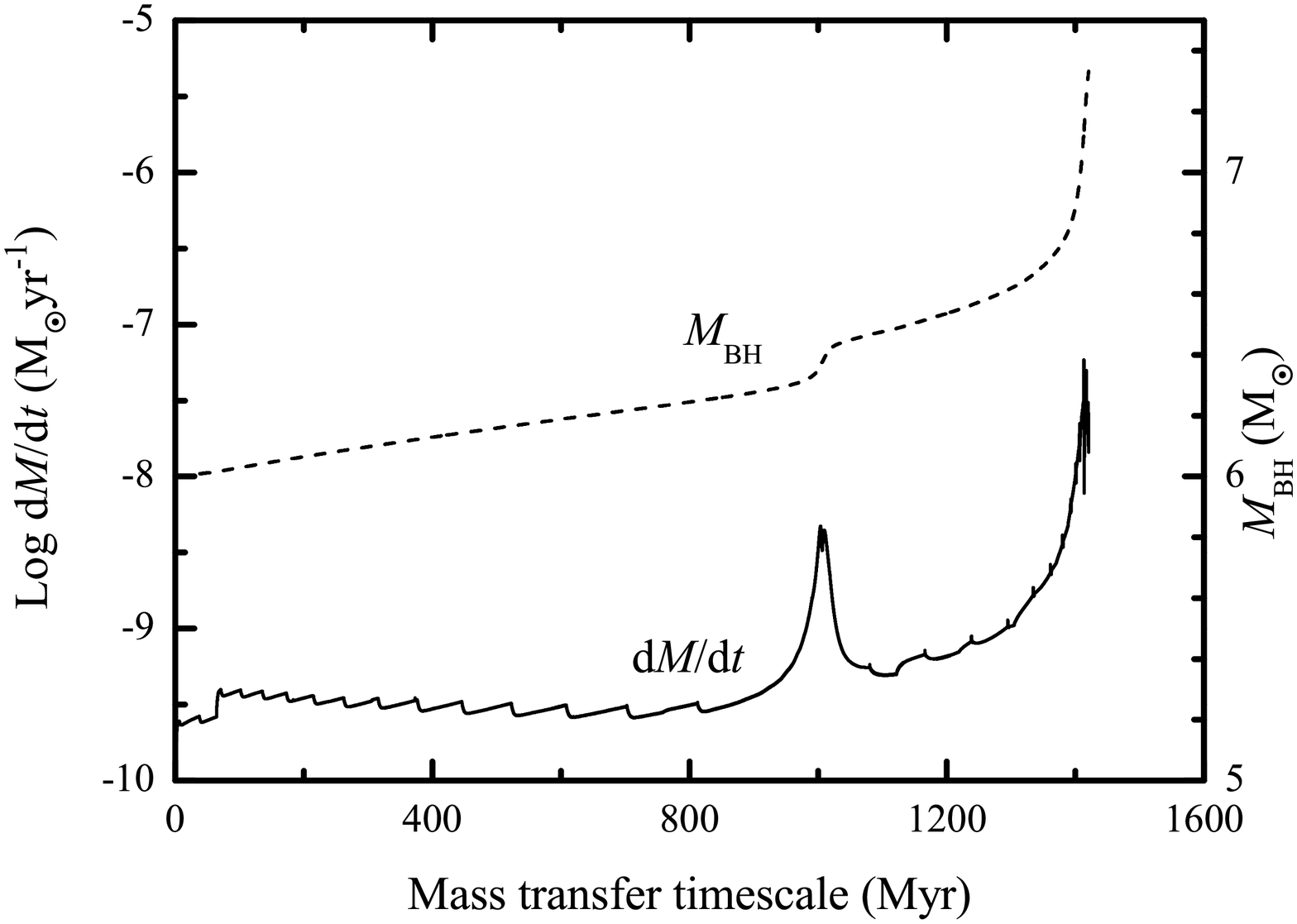}
\includegraphics[angle=0,width=9cm]{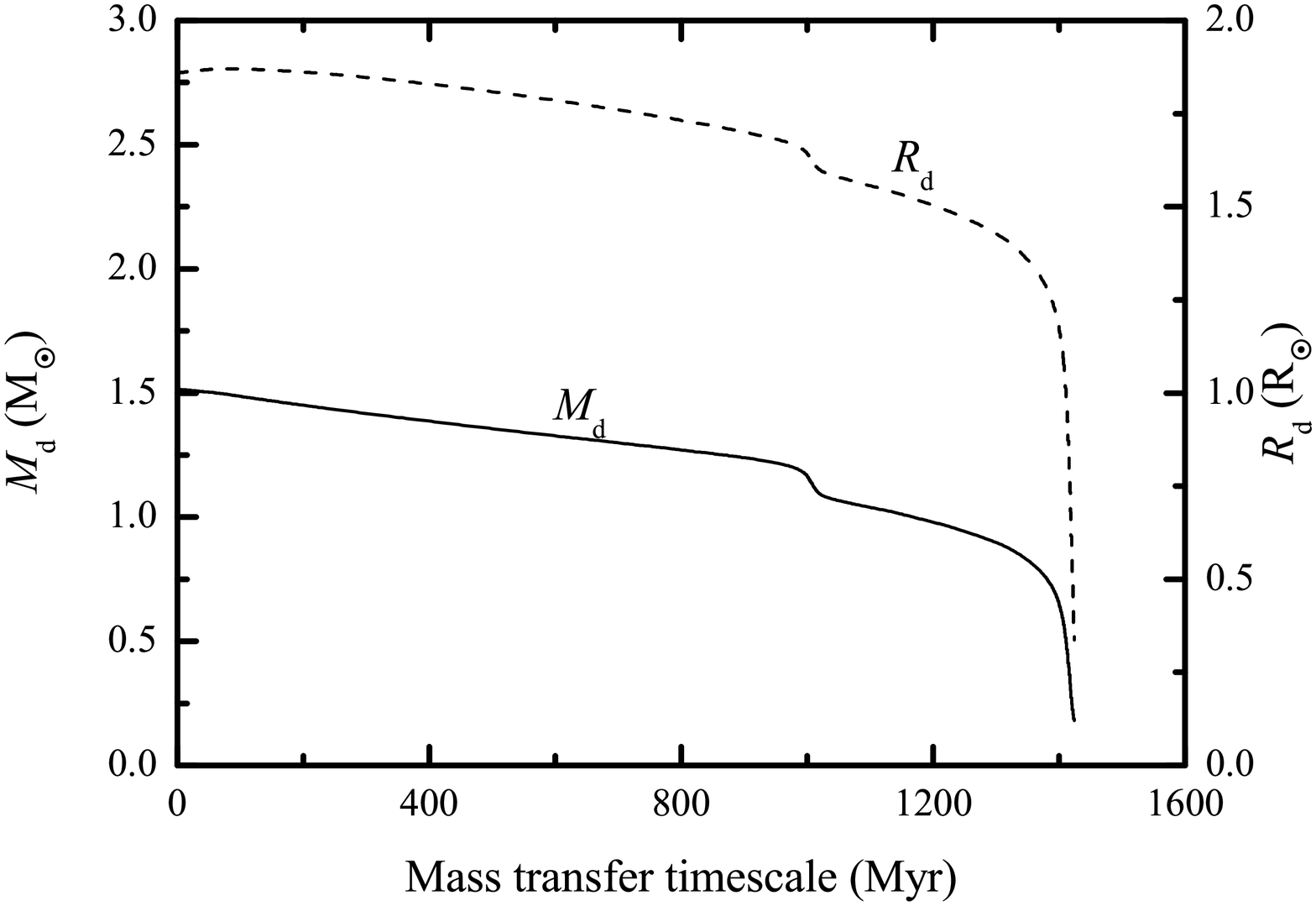}
\caption{Evolutionary tracks of BHXB
with an initial orbital period of 0.84 d, which consists of a
donor star of $M_{\rm d} = 1.5~ \rm M_\odot$ and a black hole of
$M_{\rm bh} = 6.0~ \rm M_\odot$. The solid and dashed curves
represent the evolution of the mass transfer rate and BH mass in
the top panel, and the donor star mass and donor star radius in
the bottom panel, respectively.  } \label{Fig1}
\end{figure}

\begin{figure} \centering
\includegraphics[angle=0,width=9cm]{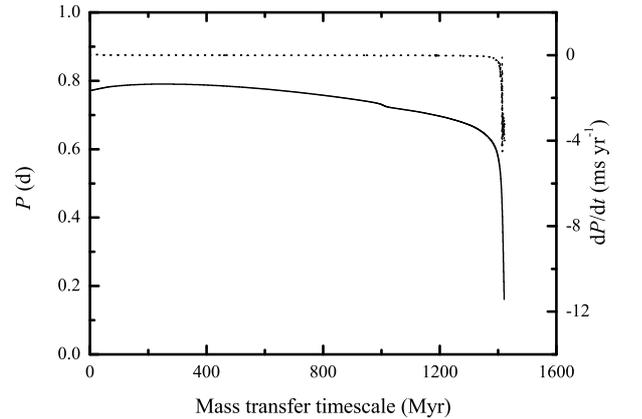}
\caption{Evolutionary tracks of BHXB with an
initial orbital period of 0.84 d, which consists of a donor star
of $M_{\rm d} = 1.5~ \rm M_\odot$ and a black hole of $M_{\rm bh}
= 6.0~ \rm M_\odot$. The solid and dotted curves represent the
evolution of the orbital period and orbital period derivative,
respectively. } \label{Fig1}
\end{figure}

\begin{acknowledgements}
We are grateful to the anonymous referee for helpful comments.
This work was partly supported by the National
Science Foundation of China (under grant number 11133001,
11173018, and 11333004), and Innovation Scientists and Technicians
Troop Construction Projects of Henan Province, China.
\end{acknowledgements}


\begin{thebibliography}{}
\bibitem[Casares 2007]{casa07} Casares, J. 2007, in IAU Symp.
238, Black Holes from Stars to Galaxies -- Across the Range of
Masses, ed. V. Karas \& G. Matt (Cambridge: Cambridge Univ.
Press), 3
\bibitem[Chen \& Li (2006)]{chen06} Chen, W. -C., \& Li, X. -D. 2006, MNRAS, 373, 305
\bibitem[Chen et al. 2006]{chen06b} Chen, W.-C., Li, X.-D., \& Qian, S.-B. 2006, ApJ, 649, 973
\bibitem[Cordes \& Shannon 2008]{cord08} Cordes, J. M., \& Shannon, R. M. 2008, ApJ, 682, 1152
\bibitem[Dervi\c{s}o\v{g}lu et al. 2010]{devi10} Dervi\c{s}o\v{g}lu, A., Tout, C. A., \& Ibano\v{g}lu, C. 2010, MNRAS, 406, 1071
\bibitem[Dubus et al. 2004]{dubu04} Dubus, G., Campbell, R., Kern, B., Taam, R. E., \& Spruit, H. C.
2004, MNRAS, 349, 869
\bibitem[Eggleton (1971)]{egg71} Eggleton, P. P. 1971, MNRAS, 151, 351
\bibitem[Eggleton (1972)]{egg72} Eggleton, P. P. 1972, MNRAS, 156, 361
\bibitem[Eggleton (1973)]{egg73} Eggleton, P. P. 1973, MNRAS, 163, 279
\bibitem[Fragos et al. (2009)]{frag09} Fragos, T., Willems, B., Kalogera, V., Ivanova, N., Rockefeller,
G., Fryer, C. L., \& Young, P. A. 2009, ApJ, 697, 1057
\bibitem[Gallo et al. (2007)]{gall07} Gallo, E., et al. 2007, ApJ, 670, 600
\bibitem[Gonz\'{a}lez Hern\'{a}ndez et al. (2011)]{gonz11} Gonz\'{a}lez Hern\'{a}ndez,  J. I., Casares, J., Rebolo, R., Israelian,
G., Filippenko, A. V., \& Chornock, R. 2011, ApJ, 738, 95
\bibitem[Gonz\'{a}lez Hern\'{a}ndez et al. (2012)]{gonz12} Gonz\'{a}lez Hern\'{a}ndez, J. I., Rebolo, R., \& Casares, J. 2012, ApJ, 744, L25
\bibitem[Gonz\'{a}lez Hern\'{a}ndez et al. 2014]{gonz14} Gonz\'{a}lez Hern\'{a}ndez, J. I., Rebolo, R., \& Casares, J. 2014, MNRAS, 438, L21
\bibitem[Haswell et al. 2002]{hasw02} Haswell, C. A., Hynes, R. I., King, A. R., \& Schenker, K. 2002, MNRAS, 332, 928
\bibitem[Justham et al. 2006]{just06} Justham, S., Rappaport, S., \& Podsiadlowski, Ph., 2006, MNRAS, 366, 1415
\bibitem[Kawaler 1988]{kawa88} Kawaler, S. D. 1988, ApJ, 333, 236
\bibitem[King et al. (1996)]{king96} King, A. R., Kolb, U., \& Burderi, L. 1996, ApJ, 464, L127
\bibitem[Lee et al. 2002]{lee02} Lee, C. -H., Brown, G. E., \& Wijers, R. A. M. J. 2002, ApJ, 575, L996
\bibitem[McClintock \& Remillard (1986)]{mccl86} McClintock, J. E., \& Remillard, R. A. 1986, ApJ, 308, 110
\bibitem[Muno \& Mauerhan (2006)]{muno06} Muno, M. P., \& Mauerhan, J. 2006, ApJ, 648, L135
\bibitem[Podsiadlowski et al. 2003]{pods03} Podsiadlowski, Ph., Rappaport, S., \& Han, Z. 2003, MNRAS, 341, 385
\bibitem[Rappaport et al. 1983]{rapp83} Rappaport, S., Verbunt, F., \& Joss, P. C. 1983, ApJ, 275, 713
\bibitem[Remillard \& McClintock 2006]{remi06} Remillard, R. A. \& McClintock, J. E.
     2006, \araa, 44, 49
\bibitem[Ritter \& Kolb 2003]{ritt03} Ritter, H., \& Kolb, U., 2003, A\&A, 404, 301
\bibitem[Shakura \& Sunyaev 1973]{shak73} Shakura, N. I., \& Sunyaev, R. A. \ 1973, A\&A, 24, 337
\bibitem[Shao \& Li (2012)]{shao12} Shao, Y., \& Li, X. -D. 2012, ApJ, 756, 85
\bibitem[Sills et al. (2000)]{sill00} Sills, A., Pinsonneault, M. H., \& Terndrup, D. M. 2000, ApJ, 534, 335
\bibitem[Spruit \& Ritter 1983]{spru83} Spruit, H. C., \& Ritter, H. 1983, A\&A, 124, 267
\bibitem[Spruit \& Taam 2001]{spru01} Spruit, H. C., \& Taam, R. E. 2001, ApJ, 548, 900
\bibitem[Taam \& Spruit 2001]{taam01} Taam, R. E., \& Spruit, H. C. 2001, ApJ, 561, 329
\bibitem[Torres et al. (2004)]{torr04} Torres, M. A. P., Callanan, P. J., Garcia, M. R., Zhao, P.,
Laycock, S., \& Kong, A. K. H. 2004, ApJ, 612, 1026
\bibitem[Verbunt \& Zwaan 1981]{verb81} Verbunt, F., \& Zwaan, C. 1981, A\&A, 100, L7
\bibitem[Wang \& Wang 2014]{wang14} Wang, X., \& Wang, Z. 2014, ApJ, 788, 184
\bibitem[Wang et al. 2006]{wang06} Wang, Z., Chakrabarty, D., \& Kaplan, D. L. 2006, Nature, 440, 772
\bibitem[Wu et al. (2010)]{wu10} Wu, Y. X., Yu, W., Li, T. P., Maccaronae, T. J., \& Li, X. D.
2010, ApJ, 718, 620


\end{thebibliography}
\end{document}